\documentclass[twocolumn]{aastex61}

\usepackage{multirow}
\usepackage{placeins}

\shorttitle{Multi-Star Wavefront Control}
\shortauthors{Sirbu et al.}

\begin{document}


\title{Techniques for High-Contrast Imaging in Multi-Star Systems II: \\
    Multi-Star Wavefront Control}


\author{D. Sirbu}
\affil{NASA Ames Research Center, Moffett Field, Mountain View, CA, 94035}
\author{S. Thomas}
\affil{Large Synoptic Survey Telescope, Tucson, Arizona 85719}
\author{R. Belikov}
\affil{NASA Ames Research Center, Moffett Field, Mountain View, CA, 94035}
\author{E. Bendek}
\affil{NASA Ames Research Center, Moffett Field, Mountain View, CA, 94035}



\begin{abstract}
Direct imaging of exoplanets represents a challenge for astronomical instrumentation due to the high-contrast ratio and small angular separation between the host star and the faint planet. Multi-star systems pose additional challenges for coronagraphic instruments due to the diffraction and aberration leakage caused by companion stars. Consequently, many scientifically valuable multi-star systems are excluded from direct imaging target lists for exoplanet surveys and characterization missions. Multi-star wavefront control (MSWC) is a technique that uses a coronagraphic instrument's deformable mirror (DM) to create high-contrast regions in the focal plane in the presence of multiple stars. Our previous paper introduced the Super-Nyquist Wavefront Control (SNWC) technique that uses a diffraction grating to enable the DM to generate high-contrast regions beyond the nominal region correctable by the DM. These two techniques can be combined to generate high-contrast regions for multi-star systems at any angular separation. As a case study, a high-contrast wavefront control (WC) simulation that applies these techniques shows that the habitable region of the Alpha Centauri system can be imaged reaching at least $8 \times 10^{-9}$ mean contrast in 10\% broadband light in one-sided dark holes from 1.6-5.5$\lambda/D$.

\end{abstract}


\keywords{binaries: general, binaries: visual, instrumentation, adaptive optics, planetary systems, planets and satellites: detection}



\section{Introduction}

More than 3000 exoplanets have been discovered to date, using a variety of methods: transits, radial velocity, microlensing, and direct imaging.\footnote{Data from \emph{exoplanet.eu}, January 2017} The \emph{Kepler} mission and its follow-on \emph{K2} have confirmed more than 2000 of these exoplanets including planets in the terrestrial regime. Due to the rich statistical data set provided by \emph{Kepler} coupled with an understanding of its pipeline systematics,  it has been possible to estimate that the occurrence rates of exoplanets between 1 and 2 Earth radii and within 1 AU of Sun-like stars (specifically GK dwarves) appear to be relatively common \citep{burke2015}. One of the next major steps will be for direct imaging instruments to discover and characterize such exoplanets in the Sun's local neighborhood. Direct imaging of an exoplanet orbiting a single star represents a technical challenge because the brightness ratio between a Sun-like star and an Earth-sized rocky planet in the habitable zone is approximately ten orders of magnitude \citep{desmarais2002}. Additionally, the angular separations typically require resolving capacity of around 100 milliarcseconds for direct imaging surveys within 10 parsec. 

Ground-based instruments such as GPI \citep{macintosh2014}, SPHERE \citep{beuzit2008}, and SCExAO \citep{guyon2010} have already directly imaged and characterized exoplanets. These, however, have been hot Jupiters with large orbital separations from their host star. Whereas ground-based direct  imaging instruments can take advantage of larger apertures, achievable contrast is ultimately limited by the fast dynamics of atmospheric turbulence. The coronagraphic instrument planned for the upcoming WFIRST \citep{shaklan2013} space mission will achieve deeper contrasts at smaller inner working angles. A direct imaging survey of dim, rocky planets in the habitable zone of all nearby Sun-like stars will likely be achieved only with a relatively-large aperture space telescope such as the WFIRST with starshade concept, or current HabEx \citep{mennesson2016} and LUVOIR \citep{bolcar2016} concepts. In addition, it may be possible to directly image a habitable planet around Alpha Centauri with a telescope as small as 30cm \citep{belikov2015,bendek2015}, if both stars can be sufficiently suppressed. All these instruments use either a coronagraph or starshade to suppress stellar diffraction caused by the telescope's aperture. To enable high-contrast imaging, the coronagraph is typically coupled with a wavefront control system using a deformable mirror (DM) to eliminate residual speckles formed by surface roughness and reflectivity variations across telescope optics. Laboratory testbeds have demonstrated deep contrast for different coronagraph architectures, telescope apertures, and at small angular separations for single stars \citep{cady2016,kern2016,seo2016, sirbu2016}, but not yet for multi-star systems. 

Currently envisioned ground and space-based instrument target lists contain only single star systems even though the majority of Sun-like stars reside in multi-star systems. Many multi-star systems are not considered viable targets for direct imaging because, until now, instrumental approaches have provided the means to deal only with the diffraction and aberration-induced leakage produced for the on-axis star. For a multi-star system, each off-axis star introduces additional diffraction and aberration-induced leakage, and if the stars are close enough or if the companion is bright enough, this leakage can be significant. Alpha Centauri, the nearest star system to the Sun, is a prominent example of a star system not included in target lists for coronagraph instruments such as WFIRST, because the separation between the A and B components is only a few arcseconds. In addition to gravitationally bound multi-star systems, a single star may have a foreground or background companion (optical binary) that may be significant, particularly when imaging dim stars.  

The main contribution of this paper is to present, and demonstrate via simulated examples, a technique for starlight suppression in multi-star systems. We previously argued \citep{thomas2015} that wavefront control of the second star is necessary and in theory sufficient to remove light from the second star, even for instruments with a single-star coronagraph or a single starshade. Therefore, our technique is essentially a wavefront control algorithm and is compatible with almost any existing direct imaging mission concept that has a deformable mirror, without any changes to the hardware (though having a mild grating in the beam helps with wider binaries). Our algorithm is called ``Multi-Star Wavefront Control" (MSWC) and is based on using non-redundant modes on the DM to independently control both stars. In Section \ref{sect:science}, we discuss in more detail the scientific discovery possibilities enabled by this technique, while in Section \ref{sect:challenges} we illustrate the challenges that MSWC helps overcome. Section \ref{sect:mswc} describes the MSWC technique, and as an implementation example we extend the formalism of the widely-used single star electric field conjugation (EFC) algorithm to the multi-star case. In an earlier companion paper \citep{thomas2015}, the Super-Nyquist Wavefront Control (SNWC) technique was introduced. The SNWC technique is briefly reviewed in  Section \ref{sect:snmswc} showing how enabling suppression of residual speckles beyond the nominal Nyquist controllable region is possible. A simulated case study in Section \ref{sect:alphaCenCase} shows applications of the MSWC and SNMSWC algorithms to a binary system such as the Alpha Centauri system under different conditions. As a baseline scenario, we compare traditional, Single-Star Wavefront Control (SSWC) with MSWC. We also simulate the operation of MSWC in broadband light. Finally, we demonstrate the combination of these two techniques to enable Super-Nyquist Multi-Star Wavefront Control (SNMSWC) for Alpha Centauri at both small and large angular separations and with the dark hole located beyond the DM's Nyquist limit with respect to either one or both stars. Together, these algorithms represent a solution that can be used to suppress starlight in any nearby multi-star system, with any direct imaging telescope that has a deformable mirror, with little or no changes to the hardware.

\begin{table*}
\begin{center}
\caption{Sample multi-star systems within 10pc with a Sun-like primary and a close-in companion. \label{tbl-scientificTarget}}
\begin{tabular}{llllllllll}
\tableline
Target & Spectral & Dist. & \multirow{2}{*}{Vmag} & \multicolumn{2}{c}{Comp. Ang. Sep}&  Comp. & \multicolumn{2}{c}{Comp. Obs.} & Off-Axis Leakage \\
Star & Type & (pc) & &  (arcsec) &  ($\lambda/D$) & $\Delta$Vmag & Epoch & WDS Entry  & (Contrast Floor) \\
\tableline
$\alpha$ Cen A & G2V & 1.3 & 0.0 & ~4.0 & ~71.6 & \hphantom{-}1.3 & 2016 & $14396-6050$ & 6.7e-08\\
$\alpha$ Cen B & K1V & 1.3 & 1.3 & ~4.0 & ~71.6 & -1.3 & 2016 & $14396-6050$ & 7.6e-07 \\
\rule{0pt}{4ex}$\alpha$ CMi A & F5IV & 3.5 & 0.4 & ~3.8 & 68.0 & 10.4 & 2014 & $07393+0514$ &  2.1e-11\\
\rule{0pt}{4ex}70 Oph A & K0V & 5.1 & 4.1 & ~6.5 & 116.4 & \hphantom{-}2.0 & 2016 & $18055+0230$ & 1.8e-08\\
70 Oph B & K4V & 5.1 & 6.1 & ~6.5 & 116.4 & -2.0 & 2016 & $18055+0230$ & 7.9e-07\\
\rule{0pt}{4ex}36 Oph A & K2V & 5.5 & 5.1 & ~5.1 & ~91.3 & \hphantom{-}0.0 & 2016 & $17153-2636$  & 1.7e-07 \\
36 Oph B & K1V & 5.5 & 5.1 & ~5.1 & ~91.3 & \hphantom{-}0.0 & 2016 & $17153-2636$ & 1.9e-07\\
\rule{0pt}{4ex}$\eta$ Cas A & G0V & 6.0 & 3.5 & ~12.9 & ~230.9 & \hphantom{-}4.0 & 2015 & $00491+5749$ & 8.0e-10 \\
$\eta$ Cas B & K7V & 6.0 & 7.5 & ~12.9 & ~230.9 & -4.0 & 2015 & $00491+5749$ &1.2e-06 \\
\rule{0pt}{4ex}$\mu$ Cas A & K1V & 7.6 & 5.2 & ~1.1 & ~19.7 & \hphantom{-}5.4 & 2014 & $01083+5455$ &7.8e-08 \\
\rule{0pt}{4ex}p Eri A & K2V & 7.8 & 5.7 & 11.4 & 204.1 & \hphantom{-}0.2 & 2013 & $01398-5612$ & 2.1e-08 \\
 p Eri B & K2V & 7.8 & 5.9 & 11.4 & 204.1 & -0.2 & 2013 & $01398-5612$ & 2.8e-08\\
\rule{0pt}{4ex}$\mu$ Her A & G5IV & 8.3 & 3.4 & ~0.8 & ~14.3 & \hphantom{-}7.3 & 2015 & $17465+2743$ & 4.3e-08 \\
\rule{0pt}{4ex}HD 32450 A & K7V & 8.6 & 8.3 & ~0.9 & ~16.1 & \hphantom{-}2.0 & 2014 & $05025-2115$ & 2.8e-06 \\
\rule{0pt}{4ex}$\chi^1$ Ori A& G0V & 8.7 & 4.4 & ~0.5 & ~~9.0 & \hphantom{-}3.1 & 2002 & $05544+2017$ & 9.8e-06 \\
\tableline
\end{tabular}
\end{center}
\tablecomments{Last known angular separations to the companion star are shown in units of arcseconds and $\lambda/D$ (assuming $\lambda = 650$nm and $D=2.4$m), and computed contrast floor due to the off-axis leakage from the stellar companion assuming pure phase $\lambda/20$ RMS aberrations with a spectral envelope following a $1/f^3$ power law.}
\end{table*}
 
 \section{Scientific Motivation} \label{sect:science}
 
Direct imaging of exoplanets is planned via a suite of ground-based and space-based instruments. These instruments will provide the opportunity to characterize a large number of exoplanets going beyond the statistical census obtained via indirect detections. Additionally, with sufficiently advanced direct imaging, any nearby planet can be spectrally characterized, and not just the small fraction that are transiting. Thus, direct imaging is the only currently known technique that in theory enables a complete census and spectroscopic characterization of all nearby exoplanets.

The diameter of the imaging telescope aperture imposes a set of fundamental limitations on a direct imaging survey. One such limitation is due to the photon flux from the exoplanet. Rocky exoplanets in particular exhibit low flux, and therefore the target star list is limited in distance as faraway stars require long integration times. Additionally the habitable zone of faraway stars is challenging to image as angular separations decrease. As a result, the list of target stars available for a particular aperture size is limited. In addition, many close multi-star systems are by default excluded from current target lists for direct imaging observations even though these contain a majority or at least a significant fraction of potential target stars. Space-based missions have limited apertures but because they are not limited by the atmospheric turbulence, they can achieve higher contrasts. These missions therefore tend to focus the search around Sun-like stars rather than dimmer M-dwarfs. Alpha Centauri, the nearest star system, is one example of the type of systems that could be imaged with a small dedicated telescope with a 30-45 cm aperture size \citep{belikov2015}. To emphasize the importance of surveying multi-star systems in the solar neighborhood, stellar surveys have indicated that a majority of stars are part of multi-star systems \citep{abt1983}. For example, 5 out of the nearest 7 stars are located in multi-star systems. Within 10pc of the Sun, there are 70 Sun-like stars (defined as being of FGK spectral types) out of which 43 are located in known multi-star systems. The prevalence of Sun-like stars in multi-star systems holds out to 25pc \citep{raghavan2010} and beyond \citep{tokovinin2014}. Of course, some of these have separations that are wide enough to be of no concern for direct imaging missions, but a significant fraction of them typically have a companion leak that cannot be ignored, including Alpha Centauri.

Thus, a capability to directly image the circumstellar and circumbinary environments in multi-star systems could substantially increase the possible target star list for a given space telescope aperture. Table \ref{tbl-scientificTarget} summarizes a few of the best Sun-like stars targets within 10pc which are, however, located in multi-star systems and for which the companion introduces off-axis starlight leakage limiting the achievable contrast. Due to their proximity, many of these stars represent some of the best available direct imaging targets and are included in a recent catalog ranking target stars based on SNR \citep{leger2015}. Based on the indicated epoch entry in the WDS catalog \citep{wdsRef}, the angular separations between the target star and its companion are given in arcseconds and computed in equivalent units of $\lambda/D$ for $\lambda=650$nm and $D=2.4$m (representative of WFIRST). Also shown is the contrast floor due to the off-axis star leakage resulting from $\lambda/20$ RMS phase aberrations with a power spectral envelope of $1/f^{3}$. For example, 70 Ophiucchi has two components with an angular separation of 6.5 arcseconds and a $\Delta V$ magnitude difference of 2.0. High-contrast imaging around 70 Ophiucchi A would be limited at a contrast floor of $1.9 \times 10^{-7}$  due to the off-axis contribution from its close-in and bright companion 70 Ophiucchi B. 70 Ophiucchi B is a Sun-like star and a target of interest itself, but would be limited at a shallower contrast level of $7.9 \times 10^{-7}$ due to 70 Ophiucchi A's off-axis starlight leakage contribution (which is the brighter component). 36 Ophiucchi is a triple star system, with the A and B components separated by 5.1 arcseconds and with equal visual magnitudes. The C component has a negligible leakage contribution being located beyond 700 arseconds from the AB pair. Mu Cassiopeiae A is also limited at a contrast floor shallower than $10^{-10}$ even though the B companion is a dim M-dwarf because of the close-in nature of the binary system with only a 1.1 arcsec angular separation. On the other hand, Procyon (Alpha Canis Minoris) is sufficiently bright that leakage contribution from its M-dwarf companion separated by 3.8 arcseconds is negligible (contrast is limited contrast to $2.1 \times 10{-11}$).

\begin{figure*}
\centering
\includegraphics[width= 0.7 \textwidth]{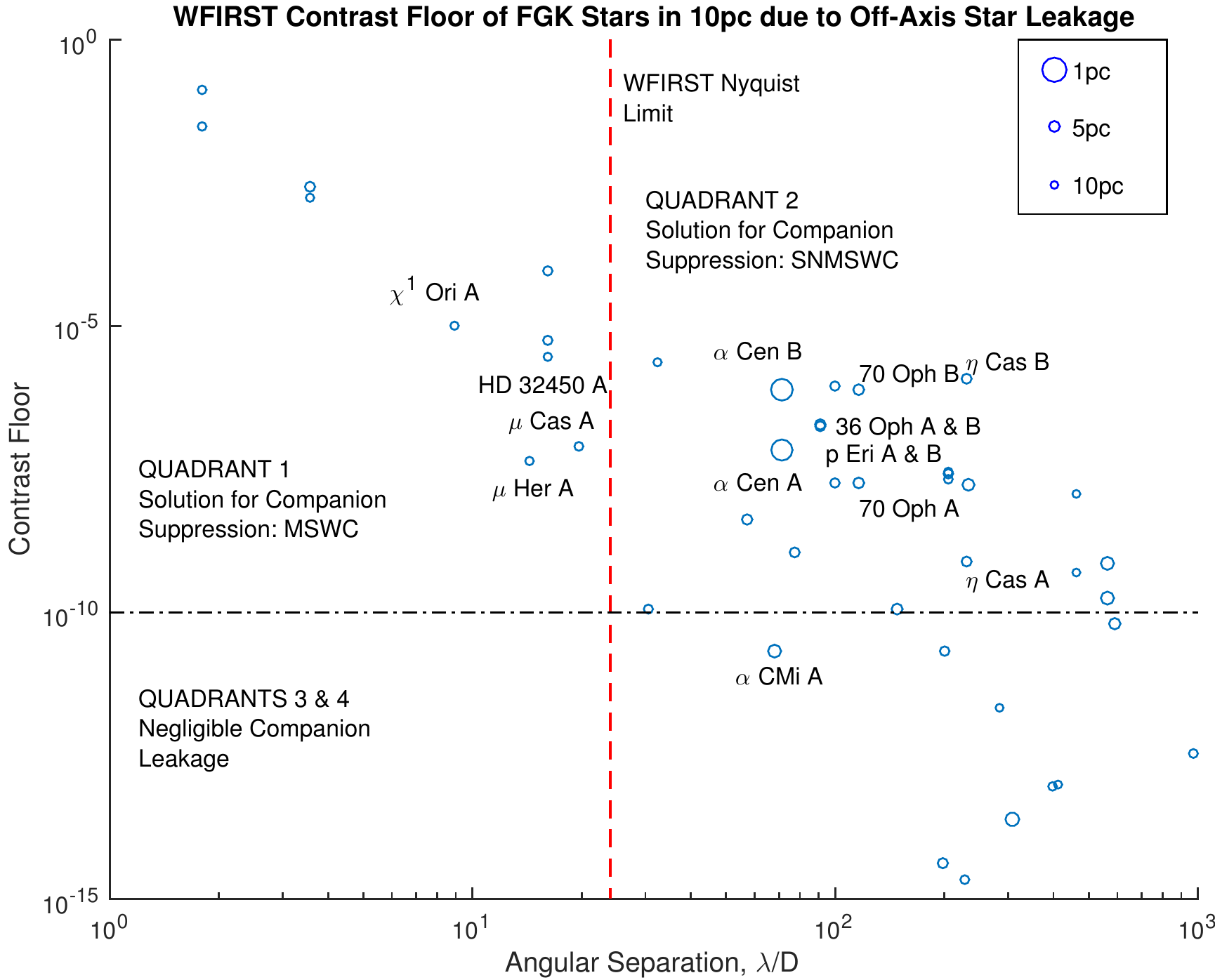}
\caption{There are 70 FGK stars within 10pc. Assuming a $D = 2.4$m primary with $\lambda/20$ RMS pure phase aberrations (with a power spectral envelope following a $1/f^{3}$ power law) and $\lambda = 650$nm, the off-axis starlight from a companion introduces a contrast floor shallower than $10^{-10}$ for 36 of these stars. Shown also is the Nyquist limit for a $48\times48$ DM, with 10 stars at Sub-Nyquist angular separations. \label{fig:contrastFloor}}
\end{figure*}

To generalize these results, in Figure \ref{fig:contrastFloor} the contrast floor induced by off-axis starlight leakage is shown for the same configuration representative of WFIRST and with all stars listed in Table \ref{tbl-scientificTarget} labeled. Out of 70 Sun-like stars within 10pc, the leakage due to an off-axis companion limits contrast at a level shallower than $10^{-10}$ for 36 stars as indicated by the horizontal dash-dot black line. These potentially scientifically interesting stars would probably be excluded from WFIRST's target list unless the companion can be suppressed (with little or no hardware changes to WFIRST). Also shown by the vertical dashed red line is the Nyquist limit of the currently baselined $48\times48$ actuator DM on WFIRST. Stars located in the bottom two quadrants (below the dash-dot black line) feature negligible leakage due to their companion and can be treated as if single stars. The companion leakage for Sub-Nyquist multi-star systems located in the top, left quadrant can be removed using only the MSWC algorithmic solution. Finally, for stars with Super-Nyquist angular separation the companion leakage can be removed with the SNMSWC algorithm combined with a diffraction grating. Note that for this figure only stars within 10pc were considered -- for stars beyond 10pc there would be additional targets in the MSWC quadrant.

There is, of course, a question whether exoplanets are expected to be present in the dynamical environment of multi-star systems. A number of studies have addressed planet formation mechanisms in binary systems concluding that circumbinary planet formation is similar to formation around a single star and that circumstellar planet formation is also possible with restrictions on orbit inclinations \citep{quintana2007, duchene2010} and maximum stable semi-major axis. Another important consideration is the long-term dynamical stability of exoplanets in circumstellar orbits. Circumstellar dynamically stable regions have been shown to exist in theory and depend on the mass ratio of the stars and their orbit eccentricity \citep{holman1999} -- all target stars listed in Table \ref{tbl-scientificTarget} have habitable zones \citep{kopparapu2013} that are (at least partly) within the dynamically stable region. Additionally, planets in multi-star systems have been confirmed with a tally of 19 circumbinary and 34 circumstellar exoplanets around mostly binary stars but also including two triple star systems \citep{bechter2014}; a recent study of Kepler candidates has shown, however, that their occurrence rates may be lower around multi-star than single-star systems \citep{kraus2016}. Thus, in addition to detecting more planets, direct imaging in the circumstellar environment would provide additional data to inform planet formation theories for multi-star systems.

\begin{table*}[t]
\begin{center}
\caption{Summary of multi-star imaging scenarios and applicability of SNWC, MSWC, and SNMSWC.  \label{tbl-multiStarScenarios}}
\begin{tabular}{cccc}
\tableline\tableline
\multicolumn{2}{c}{Multi-Star Imaging Scenario} & \multicolumn{2}{c}{WC Solution for Companion Suppression} \\
On-axis blocker & Off-axis blocker & Ang. Sep. \textless $\frac{N}{2}\lambda/D$ & Ang. Sep. \textgreater$\frac{N}{2}\lambda/D$ \\
\tableline 
Coronagraph & None (WC only)\tablenotemark{a} & {\bf MSWC} & {\bf SNMSWC} \\
Coronagraph & 2nd Coronagraph\tablenotemark{b} & {\bf MSWC} & {\bf SNMSWC}\\
Coronagraph & Starshade\tablenotemark{c} & SSWC & SSWC \\
Starshade & None (WC only)\tablenotemark{d} & SSWC & {\bf SNWC} \\
Starshade & Coronagraph\tablenotemark{e} & SSWC & {\bf SNWC}\\
Starshade & 2nd Starshade\tablenotemark{f} & N/A & N/A \\
\tableline 
\end{tabular} 
\end{center}
\tablenotetext{a}{Existing missions are already capable of MSWC with no hardware modifications. This can be extended to SNMSWC if there is quilting on the DM or a mild grating in the pupil plane.}
\tablenotetext{b}{The second (off-axis) coronagraph is not necessary for a well-baffled telescope, but may relax the stroke requirement on the DM for close stars.}
\tablenotetext{c}{Adding an off-axis starshade effectively reduces binaries to the single-star wavefront control (SSWC) suppression problem.}
\tablenotetext{d}{Adding a deformable mirror (without a coronagraph) to a starshade mission enables multi-star suppression.}
\tablenotetext{e}{The off-axis coronagraph is not necessary for a well-baffled telescope, but may relax the stroke requirement on the DM for close stars.}
\tablenotetext{f}{Adding a second starshade means that no active wavefront control system is needed.}
\tablecomments{$N$ is number of actuators across one side of the DM, WC is Wavefront Control, SSWC is Single-Star Wavefront Control, MSWC is Multi-Star Wavefront Control, SNWC is Super-Nyquist Wavefront Control, SNMSWC is Super-Nyquist Multi-Star Wavefront Control.}
\end{table*}

In Section \ref{sect:alphaCenCase} of this paper, we simulate a direct imaging scenario for the Alpha Centauri system, which at 1.3pc away from the Sun is a compelling science target. Due to its proximity, the system could be imaged in at least 3 times higher spatial and spectral resolution (in the photon-noise limited regime) than any other star, or have at least a 3 times improvement in the signal-to-noise ratio \citep{belikov2015,leger2015}. Recently, a planet candidate with a minimum 1.3 Earth mass was discovered around Proxima Centauri \citep{escude2016}, an M-dwarf star that is far (although likely dynamically bound) from Alpha Centauri A and B \citep{laughlin2006}. The Alpha Centauri stars feature an eccentric orbit with an 80 year period leading to stellar separations varying between 11-36 AU, and whose dynamically stable region has been studied \citep{wiegert1997,quintana2002, quarles2016}. The stability limits for the semi-major axes for a circumstellar exoplanet's orbit about Alpha Centauri A or B are $2.78\pm0.65$ AU and $2.49\pm0.71$ AU respectively. Additionally, the likeliest inclination of the planetary orbits is in the plane of the binary with a maximum stable inclination of approximately $60^\circ$. These stability regions for the Alpha Centauri system include the full habitable zones for both stars and are computed following the method in \citet{kopparapu2013}. Alpha Centauri A is a G-type star with a habitable zone spanning 0.9-2.2AU, while Alpha Centauri B is a K-type star with a habitable zone spanning 0.6-1.3AU.

Finally, the MSWC technique to directly image multi-star systems is not limited to dynamically-bound star systems. Indeed, a combination of foreground and background stars can form an optical multi-star system which would exhibit the same imaging challenges. The technique could reduce epoch restrictions on follow-up observations imposed by nearby background stars and the proper motion of a potential target star, and would enable imaging when background stars are not identifiable a priori (for example obscured by the diffraction halo of the target star). Out of 70 FGK stars within 10pc, 52 stars have a recorded companion (including optical companions) including 43 stars located in multi-star systems.

\section{Multi-Star Imaging Challenges}  \label{sect:challenges}

Direct imaging of multi-star systems is more challenging in comparison to a single star system because of light coming from off-axis star(s) in addition to light from the central star. The off-axis starlight must be suppressed over the same region as the on-axis, central star to create a multi-star dark hole. An option that has been proposed is to design a coronagraph that suppresses both the on-axis and off-axis stars \citep{cady2011}. However, the off-axis coronagraph would only block  diffraction from the off-axis star and aberration-induced speckles would still be present. Speckle removal requires a wavefront control system \citep{thomas2015}. Thus, for a coronagraph to create a high-contrast region and image the circumstellar region of a multi-star system, a wavefront control solution is necessary. Depending on final contrast level requirements the wavefront control system could also be sufficient to remove the known diffracted light (e.g. Airy rings) coming from the off-axis star, obviating the need for a two-star coronagraph to suppress diffraction from the off-axis star. An off-axis coronagraph could, however, be helpful in reducing the stroke requirements on the DM for close stars. One exception to this discussion would be the case of a starshade blocking the off-axis starlight -- since the occulter is external to the telescope, speckles from the off-axis star are no longer a limiting factor. The different multi-star imaging scenarios and usage of wavefront control techniques for each case at different multi-star separations is summarized in Table \ref{tbl-multiStarScenarios}.

\begin{figure*}[ht!]
\includegraphics[width=\textwidth]{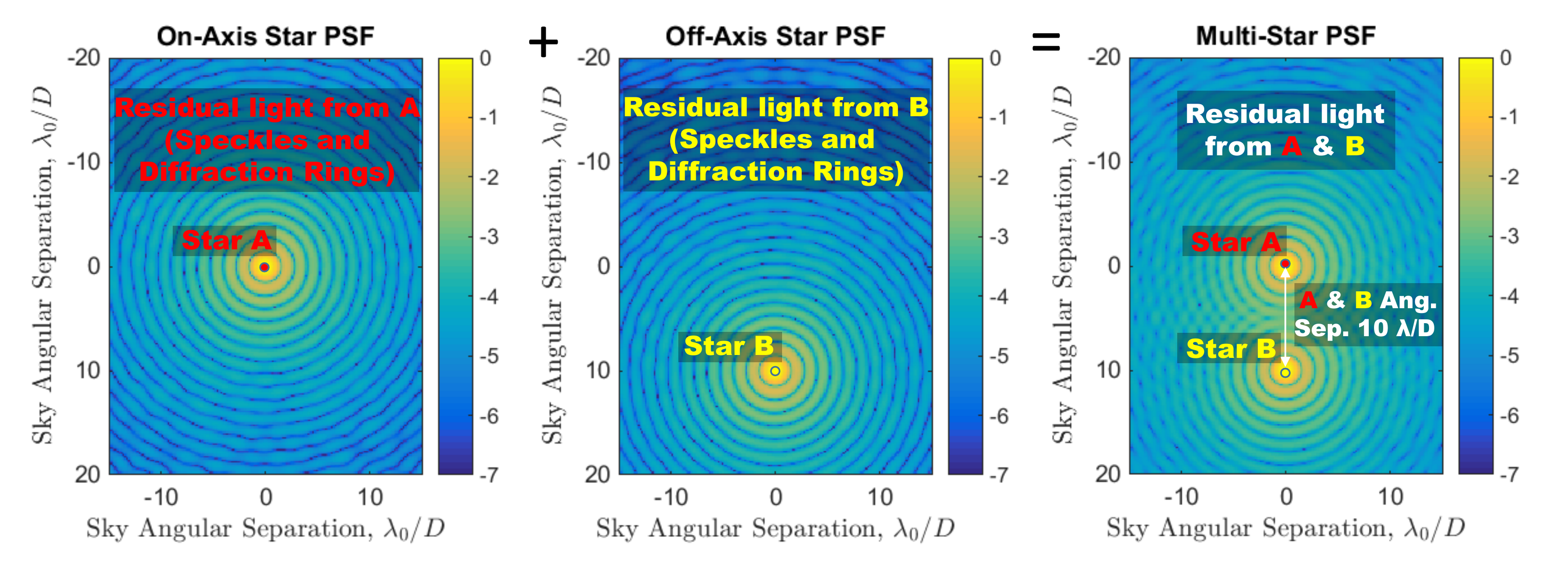}
\caption{The multi-star PSF is the incoherent sum of the PSFs for each individual star. Thus, an off-axis star limits the achievable contrast in the dark hole to the level set by its own diffraction rings and aberration-induced speckles even if a DM is used to create a dark hole around the on-axis star.  \label{fig:multiStarPSF}}
\end{figure*}

To illustrate the challenge of creating a high-contrast region for a multi-star system, refer to Figure \ref{fig:multiStarPSF} which represents a simple unobscured circular pupil with no coronagraph featuring an on-axis and an off-axis star with 10$\lambda/D$ angular separation between them. Each star creates its own point spread function (PSF) in the telescope's focal plane, which are shown independently in the left and center panes of Figure \ref{fig:multiStarPSF}. The intensity of the combined multi-star PSF shown in the corresponding right pane is formed by the incoherent addition of the individual stellar PSFs. 

A separate challenge may arise due to a wide angular separation between the two stars. The particular restriction here is imposed by the spatial bandwidth available on the DM. Traditionally, the outer working angle of the wavefront control system is given by the maximum controllable frequency of the DM (its Nyquist limit). If the desired dark hole is at an angular separation with respect to each of the stars that is within the Nyquist limit then a feasible region of high-contrast can be found that simultaneously suppresses both stars. However, for stars with wider angular separations or a larger telescope aperture it is likely that the dark hole is located beyond the controllable Nyquist-limit for one of the stars. In this case, MSWC can be combined with the SNWC technique \citep{thomas2015} that allows using higher quilting orders introduced by a diffraction grating to replicate the PSF and extend the controllable spatial frequencies to the Super-Nyquist regime (for example, creating a dark zone at 100 $\lambda/D$ with a 32$\times$32 DM). As part of the specific case study in this paper, we will demonstrate how SNWC can be combined with MSWC to gain complete coverage of the dynamically stable region in which potential circumstellar exoplanets could exist in the Alpha Centauri system.

\section{Multi-Star Wavefront Control} \label{sect:mswc}

Several wavefront control techniques have been developed to eliminate residual diffraction leakage and speckles due to aberrations in the optical train for a single on-axis star.  These techniques include Electric Field Conjugation \citep{giveon2007}, Stroke Minimization \citep{pueyo2011}, and traditional Speckle Nulling. We discuss how these wavefront control techniques can be adapted for the case of a binary star system for which mutually incoherent speckles from each star overlap spatially and thus must be simultaneously controlled.

\begin{figure*}[ht!]
\includegraphics[width=\textwidth]{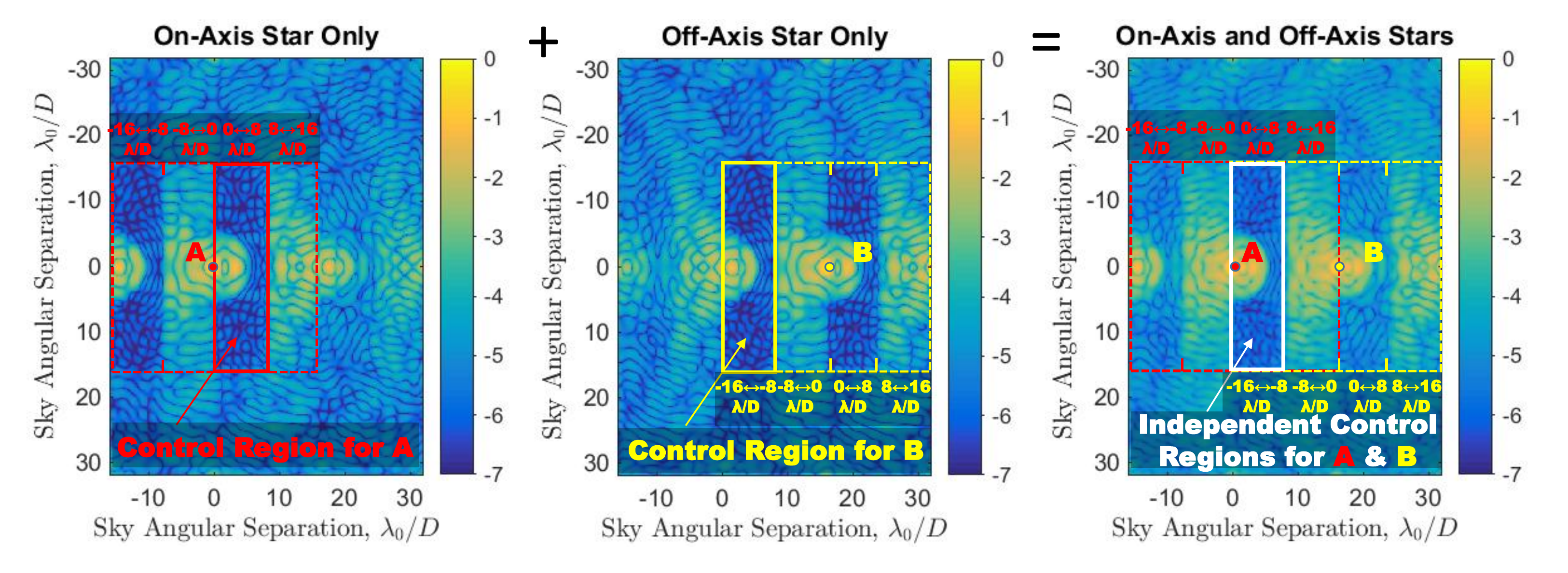}
\caption{Illustrated simulation of MSWC technique for an unobscured telescope pupil for a binary system with 16$\lambda/D$ angular separation between Stars A and B. Shown are the controllable regions with a 32 $\times$ 32 actuator DM for (\emph{Left}) the on-axis star (A) and (\emph{Center}) the off-axis star (B). (\emph{Right}) The resulting multi-star feasible dark zone using non-redundant DM modes creating independent control regions for A and B.\label{fig:multiStarWFC}}
\end{figure*}

As an illustrative example, consider Figure \ref{fig:multiStarWFC} where two stars are imaged with an unobscured pupil, no coronagraph, and separated by 16$\lambda/D$ (Nyquist separation for a DM with 32 $\times$ 32 actuators). Suppose that the desired dark hole is to be generated within 0-8$\lambda/D$ near the on-axis star and between the two stars. This dark hole could be generated for the on-axis star by applying a linear combination of sinusoidal modes with 0-8 cycles per aperture (cpa). Conversely, the same dark hole for the off-axis star would be located between 8-16$\lambda/D$ (middle panel). Therefore, the dark hole could be generated for the off-axis star using a linear combination of sinusoidal modes of 8-16cpa. These modes are independent and can therefore be used to independently generate dark holes for each star. An outstanding issue is that higher order (nonlinear) effects generate residual speckles; however, these are second-order effects that can be eliminated iteratively when operated in a closed-loop. The only requirement for this is to iterate sequentially between stars (for speckle-nulling) or to perform the wavefront control simultaneously for both stars (for model-based EFC-like algorithms) rather than performing many iterations on one star's dark hole and then many iterations on the second star's dark hole. Figure \ref{fig:multiStarPSF} shows the generated multi-star dark hole. The same DM setting obtained to simultaneously generate the multi-star dark hole in the right pane is maintained for illustrative purposes with only the on-axis and off-axis stars in the left and center panels. Thus, each star's dark hole is shown independently and their incoherent sum shows the resulting multi-star dark-hole. Since there is no coronagraph, the MSWC dark hole in this example is limited in terms of contrast depth close-in to the central star. A cost of using MSWC for this binary star scenario is that the maximum dark hole area for MSWC is a factor of two smaller than the maximum dark hole area for a single star, limited by the total number of independent degrees of freedom on the DM. For example, in a single-star case, a 32$\times$32 DM can in principle generate a dark hole that extends from 0 to 16 $\lambda/D$ (out to dotted lines) rather than from 0-8 $\lambda/D$. However, in practice, some dark zone size is usually sacrificed in order to release DM degrees of freedom to enable digging to deeper contrast levels, as well as in broader bands. Therefore, the cost in dark zone size reduction imposed by multi-star wavefront control may be much milder than a factor of 2 in practice. Also note that although our example used a star separation of 16 $\lambda/D$ and a dark zone location and geometry shown in Figure \ref{fig:multiStarPSF}, the method works for any separation greater than ~2 $\lambda/D$ (with SNWC if the separation is large), and any non-redundant dark zone location and geometry (i.e., the DM modes from  all stars do not overlap). Multiple dark holes can in principle be created serially and stitched together to form a dark zone with an arbitrary size and location. 

So far we have shown that it is possible to suppress speckles from two stars independently, but in order to compute the DM solution we also need to estimate the speckle electric field. The problem of reconstructing the electric field for each star separately can be solved using the classical DM probe pair method \citep{giveon2011}, but with a modification in the form of modulating the region of interest for each star sequentially. In the dark hole only the speckles corresponding to each star are modulated by the DM when the corresponding spatial frequencies are applied, and the speckles from the other star appear as incoherent light to the estimator, and are ignored in exactly the same way as EFC ignores actual incoherent light. As a consequence of the non-redundancy of the dark hole location, a different set of spatial frequency modes on the DM will modulate each star. 

The classical single-star EFC algorithm \citep{giveon2007} can be easily reformulated to generate a multi-star dark hole following these principles. The final electric field in the science plane $E_f$ is related to the DM electric field by the coronagraph's optical train which is abstracted as the linear operator $\mathcal{C}$:
\begin{equation}
E_f ( u, v) = \mathcal{C} \left\{  E_\mathrm{DM} (x,y) \right\}
\end{equation}
where $E_\mathrm{DM}$ is the electric field immediately after the DM plane. Then, separating the electric field into the wavefront aberration and the DM surface and applying the linear approximation:
\begin{eqnarray}
E_f ( u, v) & = & \mathcal{C} \left\{  A e^{\alpha + i \beta} e^{i \phi} \right\} \\
& \approx & E_\mathrm{ab} + i \mathcal{C} \left\{ A \phi \right\}
\end{eqnarray}
where in the above $E_\mathrm{ab} = \mathcal{C} \left\{ A e^{\alpha + i \beta} \right\}$ is the aberrated electric field with $\alpha$ the amplitude aberration and $\beta$ the phase aberration. The aberrated electric field must be corrected by the phase $\phi$ applied across the DM surface. The region in the focal plane that constitutes the continuous dark hole with respect to the on-axis star is represented by the finite set of pixels $S = \left\{ (u,v) \right\}$, where $u$ and $v$ are coordinates in the focal plane, or equivalently, spatial frequencies in the DM plane. The linearized system response relating changes in the electric field in the science plane to individual DM actuator coefficients is given by the matrix $G$ with dimensions $n_\mathrm{im} \times n_\mathrm{act}$, where $n_\mathrm{im}$ represents the number of pixels in the dark hole and $n_\mathrm{act}$ the total number of actuators across DM. The individual DM actuator coefficients $\bar{a}$ must then satisfy the following equation to correct the aberrated electric field:
\begin{equation}
G \bar{a} = - E_\mathrm{ab}
\end{equation}
where $\bar{a}$ and $E_\mathrm{ab}$ are, respectively, the DM actuator coefficients and electric field sampled by the camera pixel lattice, both flattened into 1-D vectors. Finding a solution to this equation for the DM actuator heights $\bar{a}$ represents classical EFC for a single star. 
 
To extend for the case of MSWC, the set of pixels that forms the dark hole for each star must first be defined. We assume that we have $n$ stars located in the focal plane at coordinates $(u_{*,i},v_{*,i})$, for $i=1,\ldots,n$ respectively. Without loss of generality, let $(u_{*,1},v_{*,1}) = (0,0)$, i.e. the first star is located at the origin. We will refer to the first star as ``target" or ``on-axis", and other stars as ``off-axis". We define $S_1$ as the set of $(u,v)$ corresponding to the desired region of interest where starlight suppression is to be obtained. For each $n$th star, we also define $S_i  = \{\pm(u,v)\left|\right. (u + u_{*,i}, v + v_{*,i})\in S_1\}$. In other words, $S_i$ is the region of interest $S_1$, represented in the ``local" coordinate system of the $i$-th star (i.e. one where that star is at the origin). In order for MSWC to work, we need to impose the requirement:
\begin{equation}
\cap S_i = \O
\end{equation}
We will refer to this requirement as ``non-redundancy". To first order, DM frequencies map to focal plane coordinates $(u,v)$, so a non-redundant region of interest ensures that each DM spatial frequency corrects speckles from at most one star, which is necessary because stars are incoherent with respect to each other. In effect, the degrees of freedom on the DM are distributed among the different stars. Non-redundancy also implies that for $n$ stars, the maximum MSWC controllable area is a factor of $1/n$ of a single star's controllable area. It also implies that stars cannot be closer than 1 $\lambda/D$ (otherwise non-redundant regions of interest must have feature sizes of widths smaller than $1\lambda/D$). Fortunately, there are very few stars that have such a small separation that would be of interest to direct imaging missions because they would either be very far away or unstable habitable zones. However, in general the non-redundant regions can be very flexible in their actual shape and location, and do not have to be rectangular, concave or even fully connected.

Each $i$-th star and corresponding region $S_i$ constitutes a single-star wavefront control problem. For a non-redundant region of interest, these problems can in general be solved simultaneously because they (to first order at least) use disjoint sets of modes on the DM. For example, in the case of EFC, each $i$-th star and region of interest $S_i$ has an EFC matrix $G_i$ relating the actuator coefficients and electric field inside the region of interest. A solution that suppresses speckles simultaneously for all stars can be found by solving the simultaneous set of equations:
\begin{equation}
\left[
\begin{array}{c}
G_1 \\
\vdots\\
G_n 
\end{array}
\right] \bar{a} = - \left[
\begin{array}{c}
E_\mathrm{ab,1} \\
\vdots\\
E_\mathrm{ab,n}
\end{array} \right]
\end{equation}
Solving the above system of equations for the unknown actuator response $\bar{a}$ yields the desired DM solution in the form of actuator response coefficients. The non-redundancy requirement ensures that the system is well-behaved (far from singular) so that a solution can always be found. 

Note that this formulation is with a single DM and a monochromatic correction. The generalization to broadband is identical to the single-star case \citep{giveon2007}. The linearized system response matrices are computed at each desired correction wavelength (e.g., $G_1(\lambda_1), G_1(\lambda_2), G_1(\lambda_3)$, thus reducing the broadband problem to a set of monochromatic problems to be solved simultaneously. Again, a simultaneous solution for broadband is enabled simply by stacking the individual $G$ matrices into a larger one. For example, solving the following system for actuator coefficients yields a dark zone for 2 stars and 3 wavelengths:
\begin{equation} \label{eqn:broadband}
\left[
\begin{array}{c}
G_1 (\lambda_1) \\
G_1 (\lambda_2) \\
G_1 (\lambda_3) \\
G_2 (\lambda_1) \\
G_2 (\lambda_2) \\
G_2 (\lambda_3)
\end{array}
\right] \bar{a} = - \left[
\begin{array}{c}
E_\mathrm{ab,1}(\lambda_1) \\
E_\mathrm{ab,1}(\lambda_2) \\
E_\mathrm{ab,1}(\lambda_3) \\
E_\mathrm{ab,2}(\lambda_1) \\
E_\mathrm{ab,2}(\lambda_2) \\
E_\mathrm{ab,2}(\lambda_3)
\end{array} \right]
\end{equation}

\begin{figure*}[ht!]
\includegraphics[width=\textwidth]{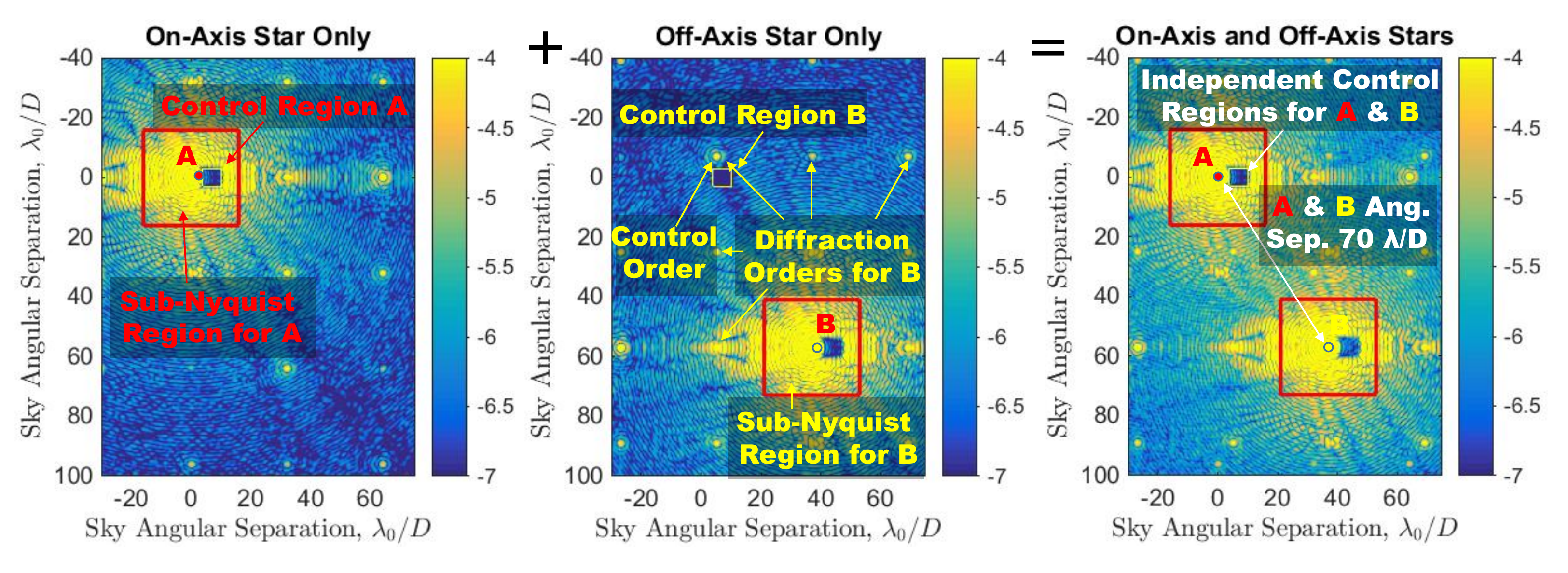}
\caption{Simulation results of MSWC for an unobscured telescope pupil for a binary system with 70$\lambda/D$ angular separation between Stars A and B. The dark hole is located outside the Nyquist limit for the off-axis star (B) requiring combination with SNWC using a diffraction grating that generates a lattice of diffraction orders. Illustrated are the independent dark holes for the (\emph{Left}) on-axis star (A), (\emph{Center}) off-axis star (B), and (\emph{Right}) combined multi-star (both on-axis and off-axis). \label{fig:superNyquistMultiStarWFC}}
\end{figure*}

\section{Super-Nyquist Multi-Star Wavefront Control} \label{sect:snmswc}

For stars at sufficiently large angular separations, MSWC alone may no longer be sufficient to create dark holes. This is because the dark regions $S_i$ for the off-axis stars can be outside the DM's controllable range (spatial Nyquist frequency). In this case, SNWC can be combined with MSWC to enable Super-Nyquist Multi-Star Wavefront Control (SNMSWC).

The combined  SNMSWC technique is illustrated in Figure \ref{fig:superNyquistMultiStarWFC} for two stars separated by a distance of $70\lambda/D$.  SNMSWC uses a diffraction grating (either in the form of already existing print-through and quilting patterns on the DM or an external grating such as a diffractive pupil \citep{bendek2013} that creates faint replicas, or diffraction orders, of the PSF at spatial frequencies beyond the Nyquist limit. These diffraction orders enable a lever to suppress starlight within their vicinity, up to the spatial Nyquist limit away \emph{from them}. Since any location in the focal plane is within the Nyquist limit of the nearest diffraction order, speckle suppression is in principle possible at any location in the focal plane (except at the diffraction order itself of course).

In the left-pane of Figure \ref{fig:superNyquistMultiStarWFC} a desired dark hole is located within the Nyquist limit for the on-axis star. Using a 32 $\times$ 32 DM's print-through pattern, a diffraction grid can be observed at regular intervals of $\pm 32\lambda/D$ across the field of view.

The central pane shows that the control region is outside the Nyquist limit with respect to the off-axis star (B). To create a dark hole in that region it will be necessary to use the nearest replica PSF from the off-axis star's diffraction orders. This PSF replica enabled modulating speckles of the off-axis star at 70$\lambda/D$ (at Super-Nyquist frequencies), and create a dark hole as shown.

Finally, in the right pane the combined multi-star dark hole is shown with both the on-axis and off-axis stars simultaneously.

As discussed in the previous subsection for MSWC, to generate the dark hole simultaneously the dark hole region location must be non-redundant. For the SNMSWC example  in Figure \ref{fig:superNyquistMultiStarWFC}, this means that the dark hole must be non-redundant with respect to the on-axis star and all off-axis PSF replicas. Thus the principles of MSWC can be combined with those of SNWC to create dark holes for multi-star systems with wider angular separations that extend beyond the Nyquist limit of the available DM.

\section{Simulated Case Study Results} \label{sect:alphaCenCase}

\begin{table*}
\begin{center}
\caption{Angular separations of objects of interest around Alpha Centauri AB and Mu Cassiopeiae A  \label{tbl-angularSep}}
\begin{tabular}{cccrrrrrr}
\tableline\tableline
\multirow{2}{*}{Target Star}  &  \multirow{2}{*}{Off-axis object} & Ang. Sep.   & \multicolumn{3}{c}{Ang. Sep (D = 0.35m)} & \multicolumn{3}{c}{Ang. Sep (D = 2.4m)} \\ 
&  & (arcsec) & $\lambda_1/D$ & $\lambda_2/D$ & $\lambda_3/D$ & $\lambda_1/D$ & $\lambda_2/D$ & $\lambda_3/D$ \\
\tableline 
$\alpha$ Cen A & B Sep. (2021) & 6 & 20.4 & {\bf 15.7} & 12.7 & 140 & 107 & 87.3 \\
$\alpha$ Cen A & B Sep. (2022) & 7 & 23.8 & {\bf 18.3} & 14.9 & 163 & 125 & 102 \\
$\alpha$  Cen A & B Sep. (2023) & 8 & 27.2 & 20.9 & 17.0 & 186 & 143 & 116 \\
$\alpha$ Cen A & Inner HZ, 0.9 AU & 0.7 & 2.36 & {\bf 1.81} & 1.47 & 16.2 & 12.4 & 10.1 \\
$\alpha$ Cen A & Outer HZ, 2.2 AU & 1.6 & 5.57 & {\bf 4.29} & 3.48 & 38.2 & 29.4 & 23.9 \\
$\alpha$ Cen B & Inner HZ, 0.6 AU & 0.4 & 1.39 & 1.07 & 0.87 & 9.55 & 7.35 & 5.97 \\
$\alpha$ Cen B & Outer HZ, 1.3 AU & 1.0 & 3.29 & 2.53 & 2.06 & 22.6 & 17.4 & 14.1 \\
\tableline
$\mu$ Cas A & B Sep. (2017) & 1.0 & 3.39 & 2.61 & 2.12 & 23.27 & {\bf 17.9} & 14.5 \\
$\mu$ Cas A & Inner HZ, 0.6 AU & 0.08 &  0.28 & 0.22 & 0.18 & 1.94 & {\bf 1.50} & 1.21 \\
$\mu$ Cas A & Outer HZ, 1.1 AU & 0.15 & 0.50 & 0.38 & 0.31 & 3.42 & {\bf 2.63} & 2.14 \\
\tableline 
\end{tabular} 
\end{center}
\tablecomments{Angular separations are converted from units of arcsec to $\lambda/D$ for both small and medium aperture telescopes (representative of ACESat and WFIRST with $D=0.35$m and $D=2.4$m respectively) at three distinct wavelengths ($\lambda_1 = 500\mathrm{nm}, \lambda_2 = 650\mathrm{nm}, \lambda_3 = 800\mathrm{nm}$).}
\end{table*}

As a simulated demonstration of MSWC, we explore a dark hole region of interest for a plausible configuration of the Alpha Centauri system. In Table \ref{tbl-angularSep}, the angular separations of the Alpha Centauri A with respect to its binary companion B are listed for three epochs. Additionally, radii of circumstellar habitable zones are computed for both Alpha Centauri A (a G-type star) and Alpha Centauri B (a K-type star) using stellar luminosity and temperature as input parameters \citep{kopparapu2013}. A dark zone spanning the habitable zone and extending out to the stability limit defines the region of interest. The habitable zone determines the necessary inner working angle close-in to the target star. Angular separations are shown in units of $\lambda/D$ for both a small telescope with an aperture of 0.35m similar to the concept in \citet{belikov2015} and for a larger telescope such as WFIRST with an aperture of 2.4m at three different representative optical wavelengths. Due to its proximity, the Alpha Centauri system is particularly well-suited to observation with a small-class telescope with angular separations within the Nyquist limit for typical DM actuator counts, while wider binaries such as 61 Cygni would require SNMSWC. Conversely, for a larger aperture the Alpha Centauri system would require SNMSWC while other potential binary target stars such as the farther away Mu Cassiopeiae fall within the Sub-Nyquist regime for typical DM actuator counts and its observation would be enabled by MSWC.

For this case study, we consider the 0.35m aperture telescope imaging the Alpha Centauri system at 650nm with a $17.5\lambda/D$ angular separation between the binary stars (corresponding to an epoch between 2021 and 2022 -- see Table \ref{tbl-angularSep}, and incidentally this separation is equally representative of Mu Cassiopeiae for a 2.4m aperture). The central star is assumed to be Alpha Centauri A which is a factor of three brighter than Alpha Centauri B. The field of view is sampled at 6 resolution elements per $\lambda/D$. The dark hole is generated with a single DM featuring 32 x 32 actuators with a working angle between 1.6$\lambda/D$ and 5.5$\lambda/D$ covering the entire habitable zone of Alpha Centauri A with the outer working angle going slightly beyond the $\approx$2.5 AU dynamical stability limit. To enable a dark hole at deep raw contrasts sufficient to directly image dim rocky planets, a coronagraph blocking diffraction from the central star is necessary because of the close inner working angle operating near the first diffraction ring. A classical phase-induced amplitude apodization (PIAA) coronagraph \citep{guyon2003} in combination with an inverse PIAA (to recover a wide field of view after blocking the central star) is used. The PIAA coronagraph is chosen because it is is well-suited for operating at small inner working angles (within a configuration similar to \citet{sirbu2016})

Notwithstanding the choice of PIAA coronagraph for this case study, the MSWC technique presented here is applicable to other coronagraph architectures as well because the generalization of EFC above is not specific to any coronagraph. Additionally, the principles of the MSWC technique can be applied to other wavefront control algorithms in addition to EFC such as classical speckle nulling.

\subsection{Baseline Case}

For this epoch and the 0.35m telescope aperture, a single-sided dark hole between $1.6-5.5\lambda/D$ is located between Alpha Centauri A and B, falling within the Sub-Nyquist controllable regime for this DM configuration for both stars. 

\begin{figure*}[ht!]
\plotone{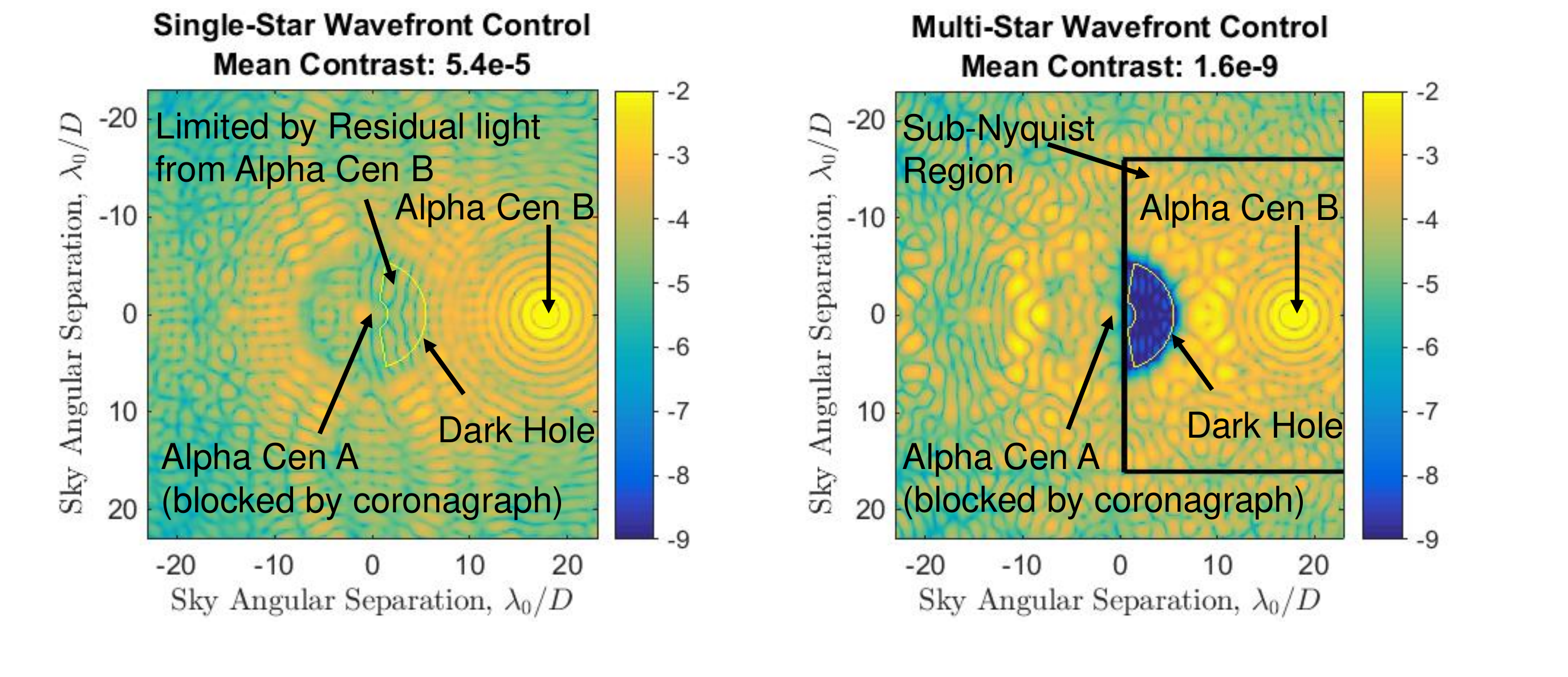}
\caption{Simulated multi-star dark hole in 650nm monochromatic light for the Alpha Centauri system with a relatively close angular separation of 17.5$\lambda/D$ between the binary stars due to a small aperture telescope ($D=0.35$m): (\emph{Left}) Using EFC around Alpha Centauri A only (the on-axis star blocked by the coronagraph), contrast in the dark hole is ultimately limited by speckles from Alpha Centauri B (the off-axis star in this configuration). (\emph{Right}) Applying EFC with MSWC, leakage from both the central star and the off-axis star can be simultaneously removed.  The Sub-Nyquist control region for Alpha Centauri B is indicated.
\label{fig:alphaCenSingleVsMultiple}}
\end{figure*}

Before the wavefront control loop is applied, at 650nm and without a coronagraph the mean contrast from the unaberrated on-axis star measures  $7.1 \times 10^{-3}$. After introduction of the coronagraph, the on-axis star's contribution is controlled by two orders of magnitude (but still dominated by diffraction) with a mean contrast contribution of $1.9 \times 10^{-5}$. The off-axis star's leakage contribution is at the same level but dominated by $\lambda/20$ phase aberrations (generated with a frequency spectral envelope following a $1/f^{3}$ power law) with a mean contrast of $1.2 \times 10^{-5}$. The combined multi-star measured contrast is the summation of the two stars' contributions at $3.1 \times 10^{-5}$.

\emph{Single-star wavefront control}. As a control case, we consider application of traditional, single-star, closed-loop EFC for the the on-axis star only starting from this initial contrast level. Alpha Centauri A is not directly visible as it is on-axis and blocked by the coronagraph, while Alpha Centauri B is off-axis and is approximately unaffected by the coronagraph after the forward PIAA optics are reversed by the inverse PIAA optics. The Nyquist region for the 32 x 32 actuator DM is shown with respect to Alpha Centauri A, with the entire dark zone at Sub-Nyquist frequencies. To first order, the DM cancels only the speckles of the on-axis star and ignores the speckles from the off-axis star because they are incoherent with respect to A. The final control region with both stars included is shown in the left pane of Figure \ref{fig:alphaCenSingleVsMultiple}, with a mean contrast of $5.4 \times 10^{-5}$ inside the dark hole. The on-axis star leakage is actually suppressed to a deep mean contrast level of $1.5 \times 10^{-10}$, but ultimately the contrast is limited by the uncontrolled off-axis star's leakage. In fact, we note that minimizing only the on-axis star leakage results in a worse contrast level inside the dark hole because the DM pattern used to suppress star A's speckles exacerbates star B's speckles in the control region. This inability to generate a dark hole by removing speckles from the on-axis star alone underlines the necessity to control the leakage of both stars simultaneously using MSWC. 

\emph{Multi-star wavefront control}. To complete the baseline scenario, we consider application of closed-loop EFC using MSWC starting from the same initial contrast level. The dark hole is created using the same physical settings as the control case. The final multi-star dark hole is shown in the right pane of Figure \ref{fig:alphaCenSingleVsMultiple}, with the Sub-Nyquist region around Alpha Centauri B indicated by the yellow outline. The dark hole for the multi-star case is  clearly visible with the mean contrast measured across is $1.9 \times 10^{-9}$, with Alpha Centauri A's contribution being $3.2 \times 10^{-10}$ and Alpha Centauri B's contribution being $1.6 \times 10^{-9}$. The Strehl Ratio of the central star is 0.92 for these DM settings, demonstrating that deep contrast can be obtained with a small impact upon the planet Strehl inside the dark hole.

\subsection{Broadband Case}

\begin{figure*}
\plotone{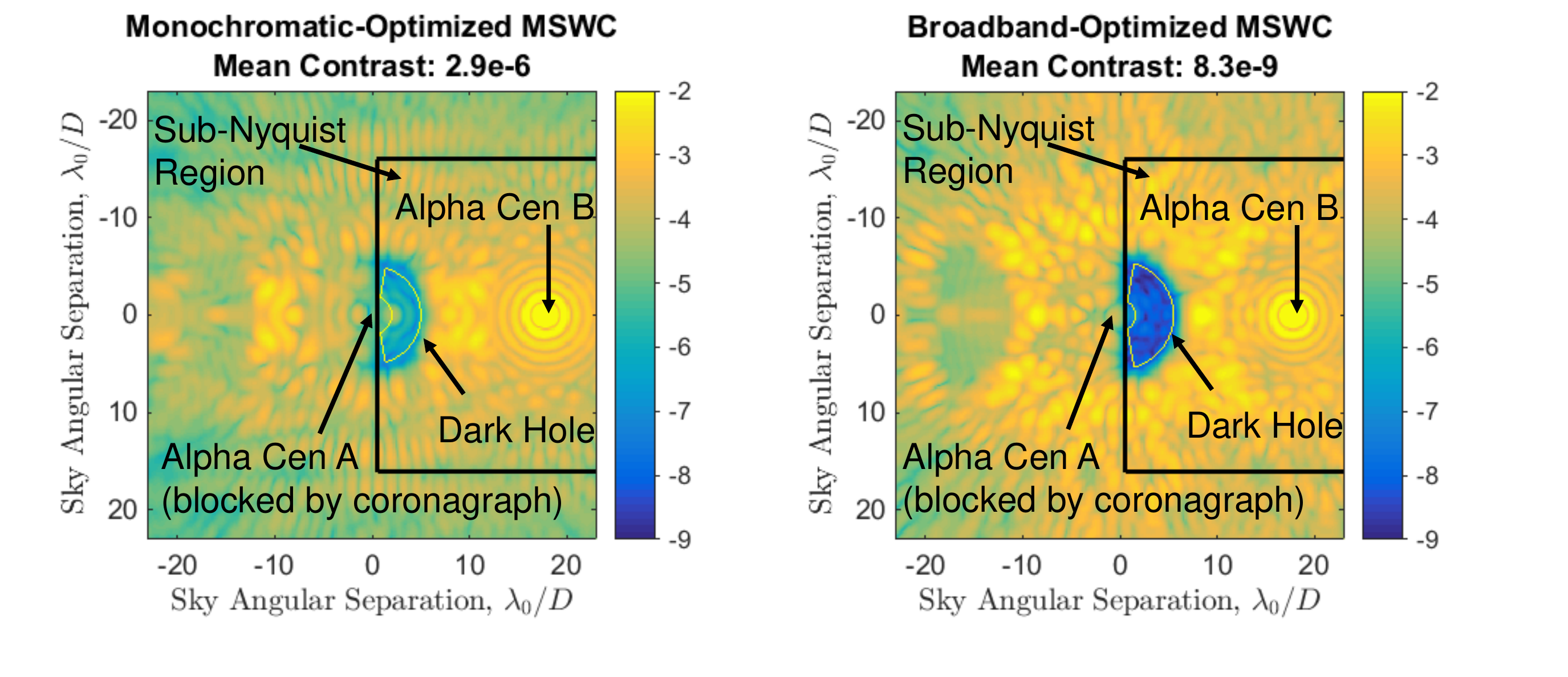}
\caption{Same scenario as Figure \ref{fig:alphaCenSingleVsMultiple} but operating in broadband light featuring a 10\% bandwidth around 650nm: (\emph{Left}) Using monochromatic-optimized MSWC wavefront control the mean contrast in the dark hole is limited by chromatic effects at $2.9 \times 10^{-6}$ (\emph{Right}) Applying broadband-optimized MSWC at three different wavelengths spanning the broadband bandwidth chromatic effects are controlled and a mean contrast of $8.3 \times 10^{-9}$ is obtained across the dark hole.
\label{fig:monoVsBroadband}}
\end{figure*}

In the baseline scenario above, we optimized DM settings for monochromatic input light. In broadband, speckles are elongated radially and can leak additional light inside the dark zone if not specifically optimized across the entire broadband bandpass. 

\emph{Monochromatic-optimized control}. This situation can be observed in the left pane of Figure \ref{fig:monoVsBroadband}, where the a MSWC DM solution is obtained in monochromatic light, but then the bandwidth is increased while keeping the DM fixed. Specifically, we used a 10\% bandpass about 650nm (generated at 1nm intervals between 617nm and 683nm). The mean contrast under these conditions degrades to $2.9 \times 10^{-6}$, nearly three orders of magnitude compared to the monochromatic case (right pane of Figure \ref{fig:alphaCenSingleVsMultiple}) for which these DM settings were specifically optimized. The chromaticity of the solution arises since the DM represents a topographic surface. As a result phase varies as $1\lambda$ across the surface height \citep{shaklan2006}. EFC generates a dark hole by minimizing aberrations in the focal plane as opposed to conjugating the phase error, so the solution is in general chromatic.

\emph{Broadband-optimized control}. To counteract chromatic effects, the dark hole is generated by simultaneously optimizing the actuator heights of the DM for three evenly-spaced wavelengths inside this bandwidth: 626nm, 650nm, and 674nm. The corresponding broadband dark hole is generated according to Equation \ref{eqn:broadband}, and the result shown in the right pane of Figure \ref{fig:alphaCenNearVsFar}. The mean measured contrast for this 10\% broadband multi-star dark hole is $8.3 \times 10^{-9}$, which represents an improvement of nearly two orders of magnitude compared to optimizing the settings monochromatically only (compare the left and right panes of Figure \ref{fig:alphaCenNearVsFar}). As larger strokes on the DM are required for broadband optimization, the Strehl Ratio of the central star is lower at 0.81. Nonetheless, this shows that MSWC can be operated in broadband with a very moderate loss of planet Strehl Ratio.

\subsection{Super-Nyquist Case}

For the epoch and telescope aperture example considered above, a single-sided dark hole located between Alpha Centauri A and B falls completely within the Sub-Nyquist controllable regime for this DM configuration for both stars. Even when stars have a relatively close angular separation there may exist potential regions of interest of the focal plane that are at Sub-Nyquist separations with respect to one star and at Super-Nyquist separations with respect to the other star. Thus a dark hole could be either partially or fully at super-Nyquist separations with respect to the off-axis star. An even more general case occurs for stars with wide angular separations such that the entire Sub-Nyquist region near the on-axis star may be super-Nyquist with respect to the off-axis star. A final case consists of a dark hole region of interest which is at super-Nyquist separations with respect to both stars. In general, all these cases are treatable with SNMSWC.

\begin{figure*}[ht!]
\plotone{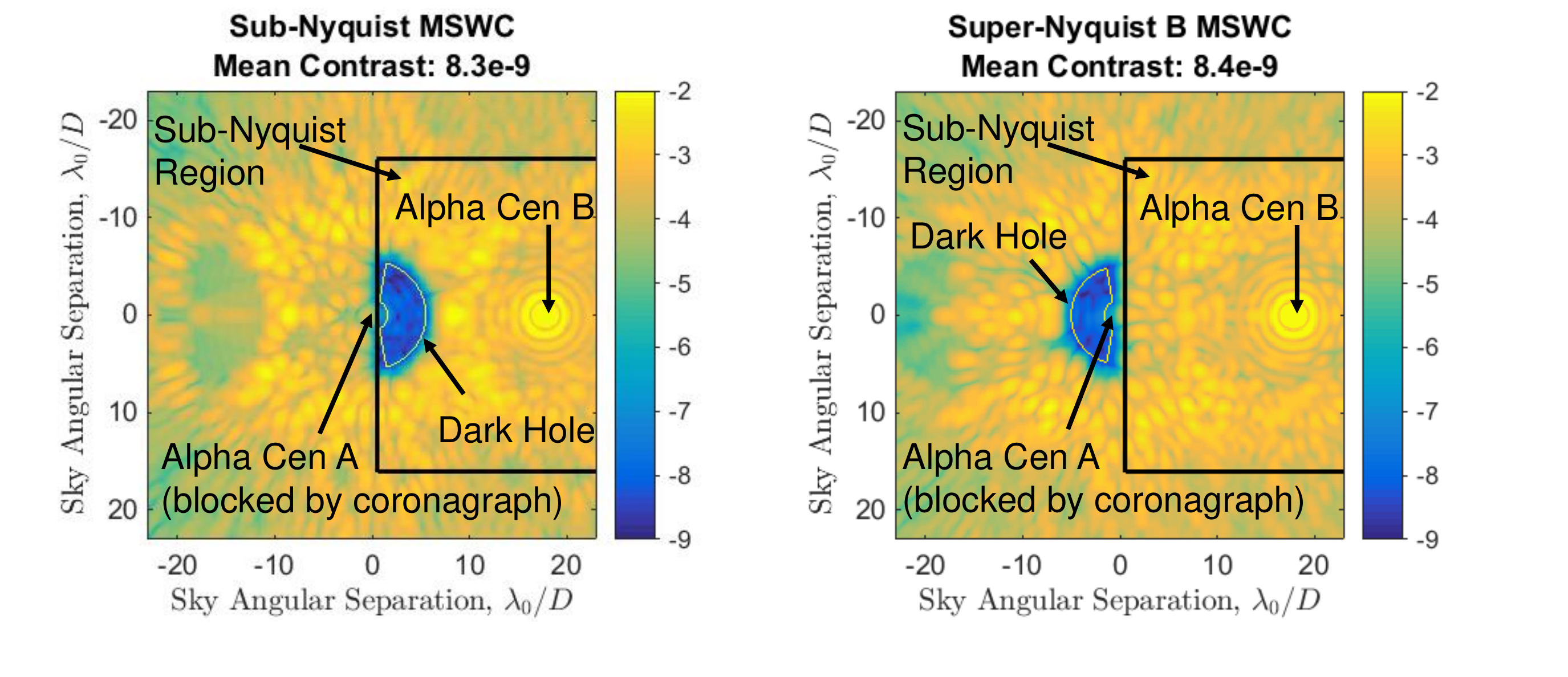}
\caption{Same broadband scenario as Figure \ref{fig:alphaCenSingleVsMultiple} but extending the controllable region to the Super-Nyquist region. The Sub-Nyquist control region for Alpha Centauri B, the off-axis star, is bounded within the indicated square. (\emph{Left}) Multi-Star Wavefront Control is used to generate a dark zone with a mean contrast of $8.3 \times 10^{-9}$ within the Sub-Nyquist region for both Alpha Centauri A and B. (\emph{Right}) Super-Nyquist Multi-Star Wavefront Control is used to generate a dark zone in the Sub-Nyquist region of Alpha Centauri A and within the Super-Nyquist region for Alpha Centauri B with a mean contrast of $8.4 \times 10^{-9}$.
\label{fig:alphaCenNearVsFar}}
\end{figure*}

\emph{Stars at small angular separations.} 
The examples above for the epoch and small telescope aperture featured Alpha Centauri A and B relatively close with a 17.5$\lambda/D$ angular separation. All these example scenarios MSWC operating at Sub-Nyquist spatial frequencies. For these cases, the single-sided dark hole is at 1.6-5.5$\lambda/D$ and located between Alpha Centauri A and B (the near-side dark hole). However, for this particular geometry a dark hole created on the opposite side of Alpha Centauri A (the far-side dark hole) lies within the Nyquist controllable limit with respect to Alpha Centauri A but outside the Nyquist controllable limit with respect to Alpha Centauri B.  Thus, this geometry requires using SNWC in combination with MSWC to generate the dark hole in the required region of interest and obtain complete coverage of the habitable zone of Alpha Centauri A for this epoch. 

To enable SNMSWC a mild diffraction grating must be introduced into the optical model \citep{thomas2015}. Here, we consider a grating already existing on many DMs, with phase arising from the DM print-through and amplitude due to etch holes. The grating model is obtained via interferometric images of a Boston Micromachines $32 \times 32$ DM and is detailed further in \citet{sirbu2016b}. The diffraction orders contain faint replica PSFs of Alpha Centauri B and are located at $\pm 32\lambda/D$ intervals away from the off-axis star. (Star A also has these replicas, but they are irrelevant.) The first diffraction order used to modulate the speckles of Alpha Centauri B in the far-side dark zone is most clearly visible in the monochromatic PSF in the left pane of Figure \ref{fig:alphaCenSingleVsMultiple}. The peak of the B PSF replica is located at -14.5$\lambda/D$ with respect to the on-axis (A)star's location.

The final dark hole generated for a 10\% bandwidth around 650nm on the far-side of Alpha Centauri B is shown in the right pane of Figure \ref{fig:alphaCenNearVsFar}, shown with respect to the Sub-Nyquist dark hole in the corresponding left pane for comparison. The dark hole has a mean contrast of $8.4 \times 10^{-9}$ and a Strehl Ratio of the central star of 0.81. These results for the far-side dark hole (super-Nyquist with respect with respect to Alpha Centauri B) are consistent with the near-side dark hole (Sub-Nyquist with respect to both Alpha Centauri A and B) and represent raw contrast without any form of post-processed speckle subtraction. Together, these results show that in principle it is possible to use a wavefront control system to generate a dark hole for a multi-star system. Additionally, these simulations have shown that the specific case of the Alpha Centauri system can be imaged with a small telescope aperture. These are raw contrasts that may be improved by further algorithm development, smaller dark zones, narrower bandwidths, use of a second coronagraph to block B, or a DM with more actuators.

\begin{figure*}[ht!]
\plotone{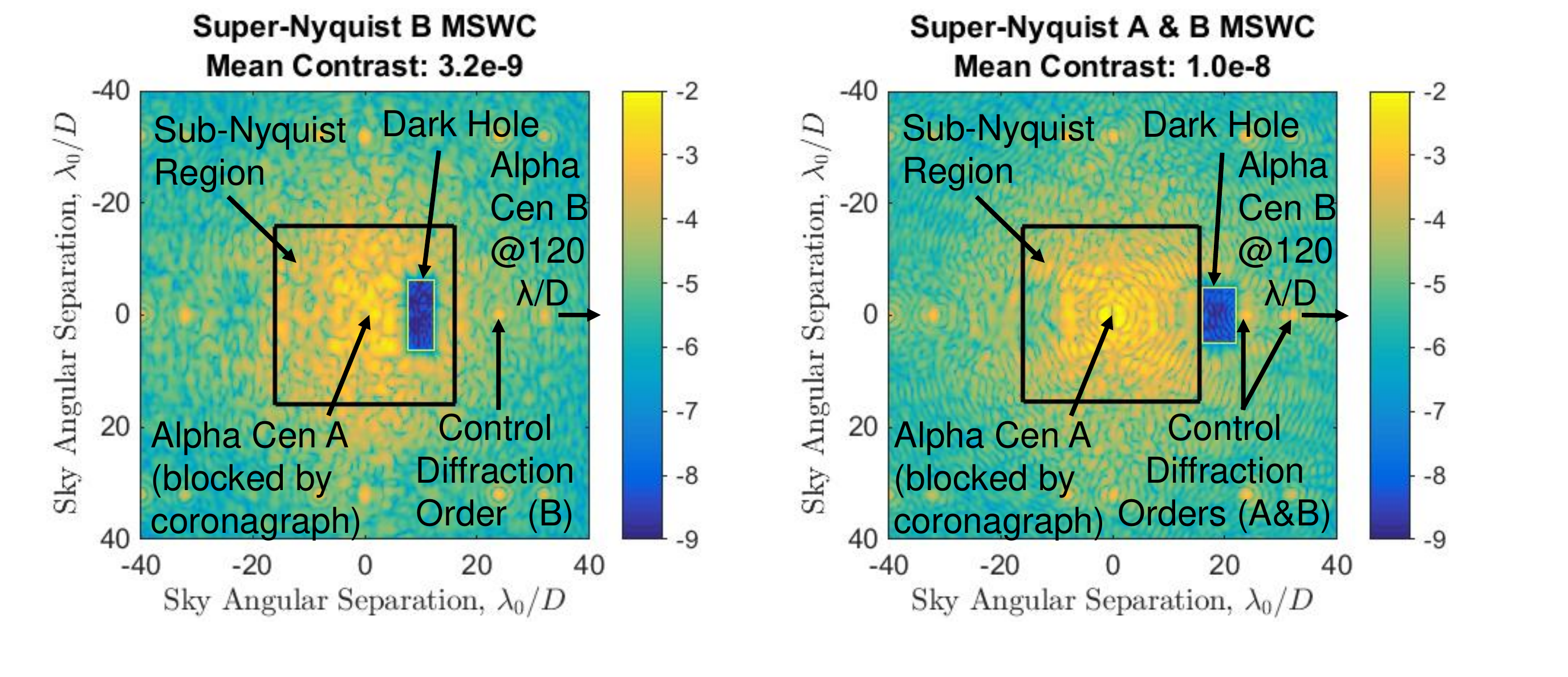}
\caption{Simulated multi-star dark hole in monochromatic light at 650nm for the Alpha Centauri system with a wide super-Nyquist angular separation of 120$\lambda/D$ between the binary stars due to a larger aperture telescope ($D=2.4$m). The Sub-Nyquist control region for Alpha Centauri A, the on-axis star, is bounded within the indicated square. (\emph{Left}) Super-Nyquist Multi-Star Wavefront Control is used to generate a dark zone with a mean contrast of $3.2 \times 10^{-9}$ with the dark zone within the Sub-Nyquist region for Alpha Centauri A and within the Super-Nyquist region with respect to Alpha Centauri B. (\emph{Right}) Super-Nyquist Multi-Star Wavefront Control is used to generate a dark zone with a mean contrast of $1.0 \times 10^{-8}$ at Super-Nyquist separations with respect to both Alpha Centauri A and B. The corresponding diffraction orders used to generate the dark zone are located at $32\lambda/D$ and $24\lambda/D$ respectively and clearly visible as indicated. 
\label{fig:largeAngularSep}}
\end{figure*}

\begin{table*}[h]
\begin{center}
\caption{Summary of performance from closed-loop wavefront control simulations of the Alpha Centauri system \label{tbl-contrastResults}}
\begin{tabular}{lllllll}
\tableline\tableline
ID\# & Stars & Simulation Description & Aberrations\tablenotemark{a} & Contrast & SR\tablenotemark{b} & Figure \\
\tableline
1 & $\alpha$ Cen A & 0.35m, monochromatic, no WC, no coronagraph & 0nm & $7.1 \times 10^{-3}$ & 1.00 & N/A \\
2 & $\alpha$ Cen A & 0.35m, monochromatic, no WC, coronagraph & 0nm & $1.9 \times 10^{-5}$ & 1.00 & N/A\\
\tableline
3 & $\alpha$ Cen A & 0.35m, monochromatic, no WC, coronagraph & 32nm & $1.9 \times 10^{-5}$ & 0.99 & N/A\\
4 & $\alpha$ Cen B & 0.35m, monochromatic, no WC, coronagraph  & 32nm & $1.2 \times 10^{-5}$ & 0.99 & N/A\\
5 & $\alpha$ Cen AB & 0.35m, monochromatic, no WC, coronagraph  & 32nm &  $3.1 \times 10^{-5}$ & 0.99 & N/A\\
\tableline
6 & $\alpha$ Cen A & 0.35m, monochromatic, after SSWC, coronagraph  & 32nm &  $1.5 \times 10^{-10}$ & 0.97 & \ref{fig:alphaCenSingleVsMultiple}, Left\\ 
7 & $\alpha$ Cen B & 0.35m, monochromatic, after SSWC, coronagraph  & 32nm &  $5.4 \times 10^{-5}$ & 0.97 & \ref{fig:alphaCenSingleVsMultiple}, Left \\ 
8 & $\alpha$ Cen AB & 0.35m, monochromatic, after SSWC, coronagraph  & 32nm &  $5.4 \times 10^{-5}$ & 0.97 & \ref{fig:alphaCenSingleVsMultiple}, Left\\ 
\tableline
9 & $\alpha$ Cen A & 0.35m, monochromatic, after MSWC, coronagraph  & 32nm &  $3.2 \times 10^{-10}$ & 0.92 &  \ref{fig:alphaCenSingleVsMultiple}, Right\\ 
10 & $\alpha$ Cen B & 0.35m, monochromatic, after MSWC, coronagraph  & 32nm &  $1.6 \times 10^{-9}$ & 0.92 &  \ref{fig:alphaCenSingleVsMultiple}, Right\\
11 & $\alpha$ Cen AB & 0.35m, monochromatic, after MSWC, coronagraph & 32nm &  $1.9 \times 10^{-9}$ & 0.92 &  \ref{fig:alphaCenSingleVsMultiple}, Right\\ 
\tableline
12 & $\alpha$ Cen AB & 0.35m, broadband\tablenotemark{c}, after MSWC, coronagraph & 32nm &  $2.9 \times 10^{-6}$ & 0.92 &  \ref{fig:monoVsBroadband}, Left\\ 
\tableline
13 & $\alpha$ Cen AB & 0.35m, broadband, before MSWC, coronagraph & 32nm &  $3.3 \times 10^{-5}$ & 0.99 & \ref{fig:monoVsBroadband}, Right\\
14 & $\alpha$ Cen AB & 0.35m, broadband\tablenotemark{d}, after MSWC, coronagraph & 32nm &  $8.3 \times 10^{-9}$ & 0.81 & \ref{fig:monoVsBroadband}, Right\\ 
\tableline
15 & $\alpha$ Cen AB & 0.35m, broadband, before SNMSWC, coronagraph & 32nm & $2.1 \times 10^{-5}$ & 0.99 & \ref{fig:alphaCenNearVsFar}, Right\\
16 & $\alpha$ Cen AB & 0.35m, broadband, after SNMSWC, coronagraph & 32nm &  $8.4 \times 10^{-9}$ & 0.81  & \ref{fig:alphaCenNearVsFar}, Right\\ 
\tableline
17 & $\alpha$ Cen AB & 2.4m, monochromatic, before SNMSWC, coronagraph & 32nm & $3.2 \times 10^{-9}$ & 0.83 & \ref{fig:largeAngularSep}, Left\\
18 & $\alpha$ Cen AB & 2.4m, monochromatic, after SNMSWC\tablenotemark{e}, coronagraph & 32nm & $3.2 \times 10^{-9}$ & 0.83 & \ref{fig:largeAngularSep}, Left\\
19 & $\alpha$ Cen AB & 2.4m, monochromatic, after SNMSWC\tablenotemark{f}, coronagraph & 32nm &  $1.0 \times 10^{-8}$ & 0.68 & \ref{fig:largeAngularSep}, Right\\ 
\tableline
\end{tabular}
\end{center}
\tablenotetext{a}{Aberrations are pure phase aberrations given in nm RMS with a spectral envelope following a $1/f^3$  power law.}
\tablenotetext{b}{The Strehl Ratio (SR) is given as a second performance metric.}
\tablenotetext{c}{The DM solution is optimized for monochromatic light (same solution as ID\# 11) but input light is broadband. The mean contrast across the dark hole is degraded due to the solution chromaticity.}
\tablenotetext{d}{The DM solution is optimized for broadband light reducing the contrast degradation due to chromaticity (compare with ID\# 12).}
\tablenotetext{e}{The dark hole is at Super-Nyquist angular separations with respect to B only.}
\tablenotetext{f}{The dark hole is at Super-Nyquist angular separations with respect to both A and B.}
\end{table*}

\emph{Stars at large angular separations.} 
Next, we consider the case of a separation between the Alpha Centauri A and B for which far super-Nyquist MSWC wavefront control is necessary. Specifically, we consider the case of a 2.4m diameter telescope aperture. Thus, Alpha Centauri A and B are separated by 120$\lambda/D$ with the potential regions of interest spanning both Sub-Nyquist and super-Nyquist separations with respect to the on-axis star (see Table \ref{tbl-angularSep}). For all regions of interest around the on-axis star, the off-axis star is at super-Nyquist separations since it is located at a wide angular separation from the on-axis star. 

Applying SNMSWC with the same DM grating model, we create a single-sided dark hole box located between $\left[7,12\right]\lambda/D$ along the separation axis between the two stars and $\left[-6,6\right]\lambda/D$ across the separation axis in monochromatic light at 650nm. This is shown in the left pane of Figure \ref{fig:largeAngularSep}. The mean contrast achieved is $3.2\times 10^{-9}$ with a Strehl Ratio of 0.83. Note that even though the angular separation between the stars has been substantially increased the off-axis star leakage contribution continues to be a limiting factor. The mean starting contrast measured across the region of interest due to the off-axis star is $5.5 \times 10^{-7}$ -- thus the final measured contrast is two orders of magnitude below the off-axis leakage contrast floor. 

Finally, we consider the case for which the dark hole is located at super-Nyquist separations with respect to both the on-axis and off-axis stars. For example, this is motivated by the fact that Alpha Centauri is so close that some of its habitable zone lies \emph{outside} the outer working angle of the WFIRST DM. The right pane of Figure \ref{fig:largeAngularSep} features a rectangular dark hole located between $\left[16,22\right]\lambda/D$ along the separation axis between the two stars, and $\left[-5,5\right]$ across the separation axis. Thus this region is outside both stars' Nyquist-controllable regions. SNMSWC is performed using the first diffraction order at $32\lambda/D$ for the on-axis star and the off-axis star's third-diffraction order located at $96\lambda/D$ with respect to the off-axis star and $24\lambda/D$ with respect to the on-axis star. The mean contrast measured across the dark hole is $1.0 \times 10^{-8}$ and the measured Strehl Ratio at 0.68.

All of the scenarios discussed in this section and performance results obtained are summarized in Table \ref{tbl-contrastResults}. The target star in all simulations was Alpha Centauri A but for some illustrative scenarios only Alpha Centauri A or B was considered as noted. The simulation performance metrics reported are the mean raw contrast measured across the dark hole and the planet Strehl Ratio. Corresponding figures are referenced.

\section{Conclusions}

A wavefront control technique that enables direct imaging of multi-star systems potentially more than doubles the number of target stars in direct imaging surveys. Additionally, it enables targeting of the Alpha Centauri system, which is significant because any telescope will receive at least ~10x the flux for any given planet type around Alpha Centauri than around any other Sun-like star, and resolve the planetary system in at least ~3x more angular resolution. This may enable characterization of potentially habitable worlds (if any exist around Alpha Centauri) on missions next decade, including a dedicated small telescope aperture telescope such as ACESat or possibly WFIRST if its DM has a strong enough quilting pattern.

In this paper, we have introduced Multi-Star Wavefront Control (MSWC), a wavefront control technique that can be used with existing coronagraph instruments to suppress starlight in multi-star systems, enabling direct imaging of circumstellar and circumbinary environments and planets in such systems. We have demonstrated through simulation that the MSWC technique enables, in principle, the imaging of the habitable zone of the Alpha Centauri system. Furthermore, this technique can be combined with the  Super-Nyquist Wavefront Control (SNWC) \citep{thomas2015} to enable Super-Nyquist Multi-Star Wavefront Control (SNMSWC). SNMSWC enables direct imaging in multi-star systems with a wide range of angular separations between the host stars.

A future paper will provide a laboratory-based demonstration of dark holes generated with SNWC, MSWC, and SNMSWC. Preliminary experimental results are reported in \citet{belikov2016}. 



\acknowledgments

This work was supported in part by the National Aeronautics and Space Administration's Ames Research Center, as well as the NASA Astrophysics Research and Analysis (APRA) program through solicitation NNH13ZDA001N-APRA at NASA's Science Mission Directorate. It was carried out at the NASA Ames Research Center. Any opinions, findings, and conclusions or recommendations expressed in this article are those of the authors and do not necessarily reflect the views of the National Aeronautics and Space Administration. DS was supported for part of this work by a NASA Postdoctoral Program fellowship. This research has made use of the Washington Double Star Catalog maintained at the U.S. Naval Observatory.  We thank Stuart Shaklan for helpful comments that improved the manuscript.

\end{document}